\documentclass[lettersize,magazine]{IEEEtran}
\usepackage{amsmath,amsfonts}
\usepackage{algorithmic}
\usepackage{algorithm}
\usepackage{array}
\usepackage[caption=false,font=normalsize,labelfont=sf,textfont=sf]{subfig}
\usepackage{textcomp}
\usepackage{stfloats}
\usepackage{url}
\usepackage{verbatim}
\usepackage{graphicx}
\usepackage{cite}
\usepackage{color}
\usepackage[table]{xcolor}
\usepackage{booktabs}
\usepackage{arydshln}

\begin{document}

\title{Zero-energy Devices for 6G:\\ Technical Enablers at a Glance}

\author{Onel López,~\IEEEmembership{Member,~IEEE},
Ritesh Kumar Singh,~\IEEEmembership{Member,~IEEE},
Dinh-Thuy Phan-Huy,~\IEEEmembership{Member,~IEEE},
Efstathios Katranaras,
Nafiseh Mazloum,~\IEEEmembership{Member,~IEEE}, Riku J\"antti,~\IEEEmembership{Senior Member,~IEEE}, Hamza Khan,  Osmel Rosabal,~\IEEEmembership{Graduate Student Member,~IEEE}, Pavlos Alexias, Prasoon Raghuwanshi, \IEEEmembership{Graduate Student Member,~IEEE}, David Ruiz-Guirola,~\IEEEmembership{Graduate Student Member,~IEEE}, Bikramjit Singh,  Andreas H\"oglund, Dung Pham Van,~\IEEEmembership{Senior Member,~IEEE},
Amirhossein Azarbahram,~\IEEEmembership{Graduate Student Member,~IEEE}, Jeroen Famaey,~\IEEEmembership{Senior Member,~IEEE}
\thanks{Onel López, Osmel Rosabal, Prasoon Raghuwanshi, David Ruiz-Guirola, and Amirhossein Azarbahram are with the Centre for Wireless Communications (CWC), University of Oulu, FIN-90014 Oulu, Finland. (\{onel.alcarazlopez, osmel.martinezrosabal, prasoon.raghuwanshi, david.ruizguirola, amirhossein.azarbahram\}@oulu.fi).
Ritesh Kumar Singh and Jeroen Famaey are with the University of Antwerp and imec, Belgium. (\{ritesh.singh, jeroen.famaey\}@imec.be).
Dinh-Thuy Phan-Huy is with Orange Labs, Ch\^atillon, France. (dinhthuy.phanhuy@orange.com).
Efstathios Katranaras is with Sequans Communications, France. (ekatranaras@sequans.com).
Nafiseh Mazloum is with Sony Nordic, Lund, Sweden. (nafiseh.mazloum@sony.com).
Riku J\"antti is with Aalto University, Finland. (riku.jantti@aalto.fi).
Hamza Khan and Bikramjit Singh are with Ericsson Research, Finland. (\{hamza.khan, bikramjit.b.singh\})@ericsson.com. Andreas H\"oglund and Dung Pham Van are with Ericsson Research, Sweden (\{andreas.hoglund,
dung.pham.van\}@ericsson.com). Pavlos Alexias is with WINGS ICT Solutions, Greece. (palexias@wings-ict-solutions.eu).
}
}



\maketitle

\begin{abstract}
Low-cost, resource-constrained, maintenance-free, and energy-harvesting (EH) Internet of Things (IoT) devices, referred to as zero-energy devices (ZEDs), are rapidly attracting attention from industry and academia due to their myriad of applications. To date, such devices remain primarily unsupported by modern IoT connectivity solutions due to their intrinsic fabrication, hardware, deployment, and operation limitations, while lacking clarity on their key technical enablers and prospects. Herein, we address this by discussing the main characteristics and enabling technologies of ZEDs within the next generation of mobile networks, specifically focusing on unconventional EH sources, multi-source EH, power management, energy storage solutions, manufacturing material and practices, backscattering, and low-complexity receivers. Moreover, we highlight the need for lightweight and energy-aware computing, communication, and scheduling protocols, while discussing potential approaches related to TinyML, duty cycling, and infrastructure enablers like radio frequency wireless power transfer and wake-up protocols. Challenging aspects and open research directions are identified and discussed in all the cases. Finally, we showcase an experimental ZED proof-of-concept related to ambient cellular backscattering.
\end{abstract}

\begin{IEEEkeywords}
6G, backscattering, energy harvesting,  energy-aware protocols, sustainable IoT, TinyML, zero-energy devices
\end{IEEEkeywords}

\section{Introduction}
There is a growing interest in evolving current mobile networks to support maintenance-free devices powered by ambient energy harvesting (EH) and being smaller, less complex, and longer-lasting than existing Internet of Things (IoT) devices. 
3rd Generation Partnership Project (3GPP) standardization activities are already underway, with discussions focusing on the requirements, topologies, and taxonomies of such IoT devices, referred to as ``ambient IoT devices'' \cite{3GPPRAN}. 
Key identified use cases include inventory management (e.g., automated warehousing, 
end-to-end logistics, automobile manufacturing, airport terminal, shipping port, electronically labeled shelves, automated supply chain), connected sensors (e.g., in smart homes, agriculture, and animal farms), localization of objects (e.g., finding a remote lost item and ranging), positioning (e.g., indoor positioning service and museum guide), and commands (e.g., online modification of medical instruments status, device (de)activation, elderly health care), for both indoor and outdoor usage \cite{3GPPRAN}. 
%
Such a device class is being explored also by the European Flagship project on the sixth generation (6G) of cellular systems, Hexa-X-II, while significantly expanding the scope to include devices that feature ultra-low energy consumption and demand throughout their entire lifecycle, from manufacturing to disposal, support zero waste generation, and adhere to material circularity principles \cite{hexaX2.2023}.

\begin{table}[t!]
    \caption{Some current EH-IoT implementations}
    \vspace{-1mm}
    \label{EH_IoT}
    \centering
    \begin{tabular}{p{1.76cm} p{1.55cm} p{1.32cm} p{2.6cm}}
        \toprule
         \textbf{solution} & \textbf{EH source}  & \textbf{connectivity} &  \textbf{use cases} \\
         \midrule
         Track Extreme & light & BLE, LTE-M, NB-IoT & asset tracking \\ \hdashline
         enerSENSE & indoor light & LoRa, NFC &  smart buildings \\ \hdashline
         Series S2 & light & 2G, LTE-M &   industrial monitoring, asset tracking \\ \hdashline
         Jack & light & BLE & fleet management \\ \hdashline
         EnOcean & vibration, heat, light  & EnOcean\textregistered, Bluetooth, Zigbee & smart spaces, smart homes \\ \hdashline
         IoT Pixels & radio fre- quency (RF) & Bluetooth & smart healthcare, supply chain \\
         \hdashline
         ONiO.zero & vibration, RF, heat, light  & BLE & ultra-low power applications \\ 
         \hdashline 
         
         Infinity & heat, light & BLE & machine monitoring \\ \hdashline 
         AirCord & dedicated laser & WiFi & healthcare, gaming, retail \\ \hdashline
         Cota\textregistered\ Real Wireless Power & dedicated RF & N/A & building automation, lighting control \\ 
         \bottomrule
    \end{tabular}
\end{table}

In the scientific community, the above devices are often referred to as energy-neutral or zero-energy devices (ZEDs) \cite{hexaX2.2023,Naser.2023,Mitsiou.2023,Infinite.2023,Rosabal.2023,Liao.2023,Phan.2022}, while herein we adopt the latter (most popular) notation. 
ZEDs promote sustainable (eco-friendly, accessible, and profitable) technology and will be a cornerstone in future wireless mobile networks, distinguishing them from their predecessors.
Note that existing real-world EH-IoT implementations, as those outlined in Table~\ref{EH_IoT}, still cannot fulfill the envisioned properties and requirements of the considered ZEDs, although they are a strong step in the right direction \cite{Lopez.2023}. 
%
%
%
%
%
%
Indeed, the limited or non-existent energy storage capabilities of ZEDs alone pose stringent constraints, making the realization of viable use cases challenging.
Specifically, energy storage/harvesting may constrain the instantaneous or short-time average energy consumption at the device since the energy supply may not be guaranteed all the time, while the overall energy consumption of devices with rechargeable batteries or (super)capacitors must stay ultra-low to optimize lifespan per charge cycle.

\begin{figure}
    \centering
    \includegraphics[width=0.5\textwidth]{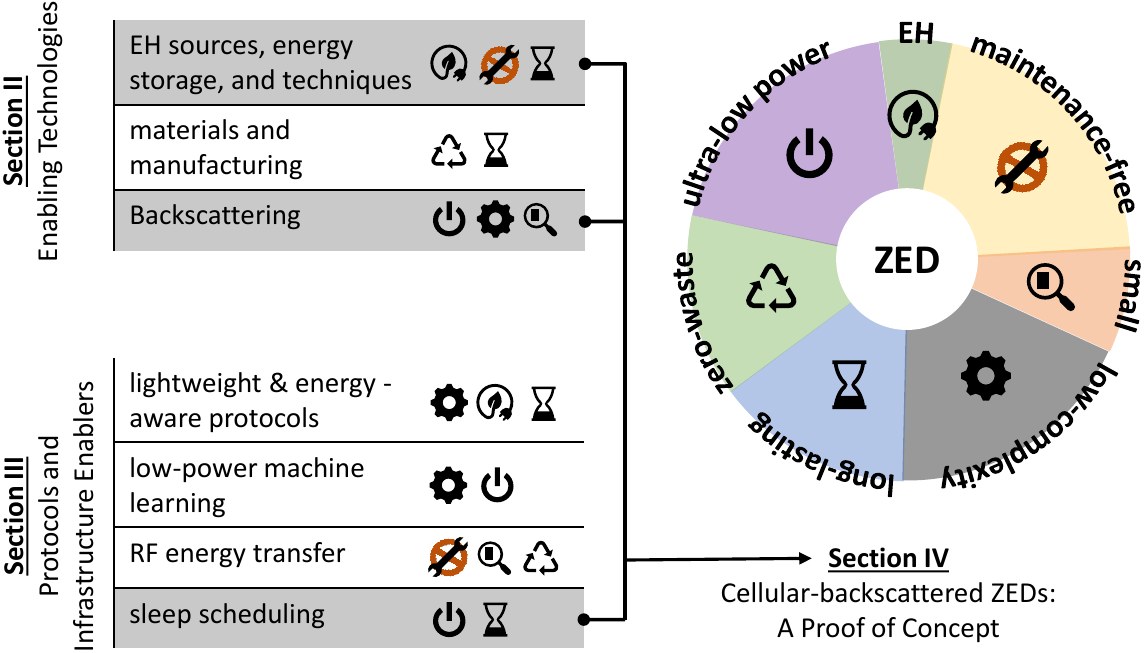}
    \caption{ZED features and enablers and corresponding section references.}
     \label{ZEDoverview}
\end{figure}

The main energy consumption sources of low-capability low-power IoT devices, and thus also ZEDs, include the connectivity module (i.e., modem); data processing (including memory reading/writing), computation, and algorithms execution; and sensing/actuation operations over/on the physical environment. Therefore, 
proper ZED designs
require 
optimized hardware components supporting ultra-low power communication, computation, sensing, and/or actuation operations. 
Note that minimizing energy usage requires the implementation of efficient management mechanisms at different levels, including device circuitry as well as network communication protocols, such as 
\emph{(i)} utilizing adaptive duty cycling and ultra-low-power sleep modes (possibly with wake-up radio implementation) at both transmitter and receiver to quickly enter and maintain deep sleeps;
\emph{(ii)} advanced low-self-discharge battery technologies for ZEDs with rechargeable energy storage;
\emph{(iii)} EH forecasting and management mechanisms; 
\emph{(iv)} adaptive energy-aware uplink/downlink transmission,
%
channel access, signaling, and scheduling; and
\emph{(v)} lightweight security mechanisms, sensing methods, and intelligence for ZEDs targeting higher-end applications. 
Finally, the corresponding manufacturing materials and processes must support the eco-friendly, low-cost, and small form-factor features of ZEDs.
Key ZED properties and technical enablers are summarized in Fig.~\ref{ZEDoverview}.

Subsets of the above technical enablers have been discussed to some extent in recent literature, e.g., \cite{3GPPRAN,hexaX2.2023,Naser.2023,Mitsiou.2023,Lopez.2023}. However, these explorations remain nascent and fragmentary, and much further in-depth research and development is needed as ZEDs usher in a  technological revolution.
In this paper, we contribute by
offering an industrial viewpoint of ZEDs toward 6G, while delving into the above technical enablers and corresponding recent advancements.
As shown in Fig.~\ref{ZEDoverview}, we focus on EH sources, energy storage solutions, manufacturing material and practices, backscattering, and low-complexity receivers in Section~\ref{sec:characteristics}. Meanwhile, we highlight the need for lightweight and energy-aware computing, communication, and scheduling protocols, while discussing potential approaches related to TinyML, duty cycling, and infrastructure enablers like RF wireless power transfer (WPT) and wake-up protocols in Section~\ref{sec:enablers}. 
It is important to note that not every technical enabler might be included in each END design, since their suitability depends on the specific characteristics and performance requirements of the ENDs and their intended applications. Meanwhile, 
in Section~\ref{sec:poc}, we showcase an experimental ZED proof-of-concept related to ambient cellular backscattering.
Challenging aspects and open research directions are identified and discussed throughout the paper, which may inspire future breakthroughs in the realm of ZEDs and set the foundation for a more sustainable and interconnected future. Finally, Section~\ref{sec:conclusion} concludes this paper.


\section{Enabling Technologies \label{sec:characteristics}} 

Herein, we explore the foundational elements and technological advancements that underpin ZEDs’ functionality. Specifically, we focus on ZEDs’ potential EH sources, energy storage solutions, and manufacturing materials and processes, while assessing their suitability and challenges. Additionally, we delve into backscattering and low-complexity receivers, key in enabling low-energy communication. 

\subsection{EH sources and techniques} 
ZEDs are powered by ambient energy sources such as light, heat, and RF signals. 
Independently of the source and corresponding EH transducers, key performance indicators (KPIs) are \emph{(i)} power density, which reveals insights on the achievable harvested energy for given transducer dimensions; \emph{(ii)} conversion efficiency, which is the percentage of the incident ambient energy converted into electricity; and \emph{(iii)} dynamic range, which provides a range of input energy levels for which the transducer conversion efficiency is above a certain value~\cite{Lopez.2023}.

Transducers' KPIs, together with their typical challenges/constraints, are crucial to preliminary assess the feasibility of EH solutions for a given application. For instance, light-based EH is usually highly efficient but might not fit scenarios where the photovoltaic cells are obstructed or shaded, as in dense forests, or hindered by dust, snow, or ice in remote areas. 
One intriguing alternative for agricultural and other environmental monitoring applications is soil thermal EH, leveraging the temperature disparity between the soil and the ambient air. 
The distinct thermal properties of soil and air contribute to a natural temperature difference, which can be harnessed using a thermoelectric generator (TEG). 
The feasibility of such an approach has been researched recently in
\cite{Puluckul.2022}, where a prototype incorporating an efficient heat transfer system is engineered. The results indicate that the TEG exhibits an average heat transfer efficiency of 30\%, demonstrating its ability to harvest energy even from temperature differences as low as 3~\textdegree C.

Key approaches to boost EH include \cite{Lopez.2023}: \emph{(i)} widening the frequency response of the transducer, e.g., multi-junction solar cells, multi-band/low-frequency vibration-based EH, and broadband/multi-band RF-EH; \emph{(ii)} capturing energy from multiple directions, e.g., omnidirectional RF-EH, multidimensional vibration-based EH, and concentrator photovoltaics; and \emph{(iii)} resorting to multi-source (hybrid) EH. 
%
%
%
%
Regarding the latter, efficient energy-combining mechanisms are required to merge energy from various sources into a storage buffer effectively. The energy from each source should be harvested simultaneously, otherwise, weaker source(s) might remain underutilized.
The latter is exactly what happens in systems using OR-ing to combine power from multiple sources through diodes. Such a technique allows current from any source to reach a load while blocking backflow, ensuring continuous power supply even if a source fails but diodes' voltage drop can prevent weaker sources from contributing effectively.
Instead, it is shown in \cite{Infinite.2023} that simultaneous EH can be accomplished by employing temporary intermediate buffers, which not only facilitate energy multiplexing but also enable the sensing of sub-milliampere harvester output currents, thereby improving the overall efficiency and reliability of the EH process. Notably, the energy combiner proposed in \cite{Infinite.2023} has 88\% efficiency independent of the number of connected sources, relying solely on the efficiency of the switching regulator. 

Although hybrid EH provides robustness and dependability guarantees, it may incur higher manufacturing costs, increase hardware complexity, and compromise the aesthetics, form factor, and lifetime (as the operating conditions may affect each transducer differently) of the devices, and thus it is not always applicable/suitable.
In general, the specific energy source(s) choice and/or merging techniques depend on their availability and strength, alongside the device's size limitations and energy requirements~\cite{Naser.2023}.

\subsection{Energy storage solutions} 
ZEDs pose a distinctive challenge by operating intermittently and gathering and storing energy in a buffer or rechargeable battery. They may initiate tasks like software execution, sensor access, and communication upon reaching a voltage threshold. However, conflicting demands arise when the power system needs to support both capacity- and temporally-constrained tasks within the same application. A larger energy buffer for capacity-constrained tasks leads to extended recharge times, hampering the reactive execution of temporally constrained tasks. Conversely, a smaller buffer favors reactive tasks but lacks the energy needed for capacity-driven operations. To enable runtime adjustments and thus accommodate different capacity requirements, \cite{Colin.2018} proposed a reconfigurable hardware energy storage mechanism. This solution allows applications to match power system characteristics to diverse task requirements, ensuring efficiency, reactivity, and adaptability compatible with various buffer types and EH setups.

\subsection{Materials and manufacturing} 
\begin{figure}[t!]
    \centering   \includegraphics[width=0.5\textwidth]{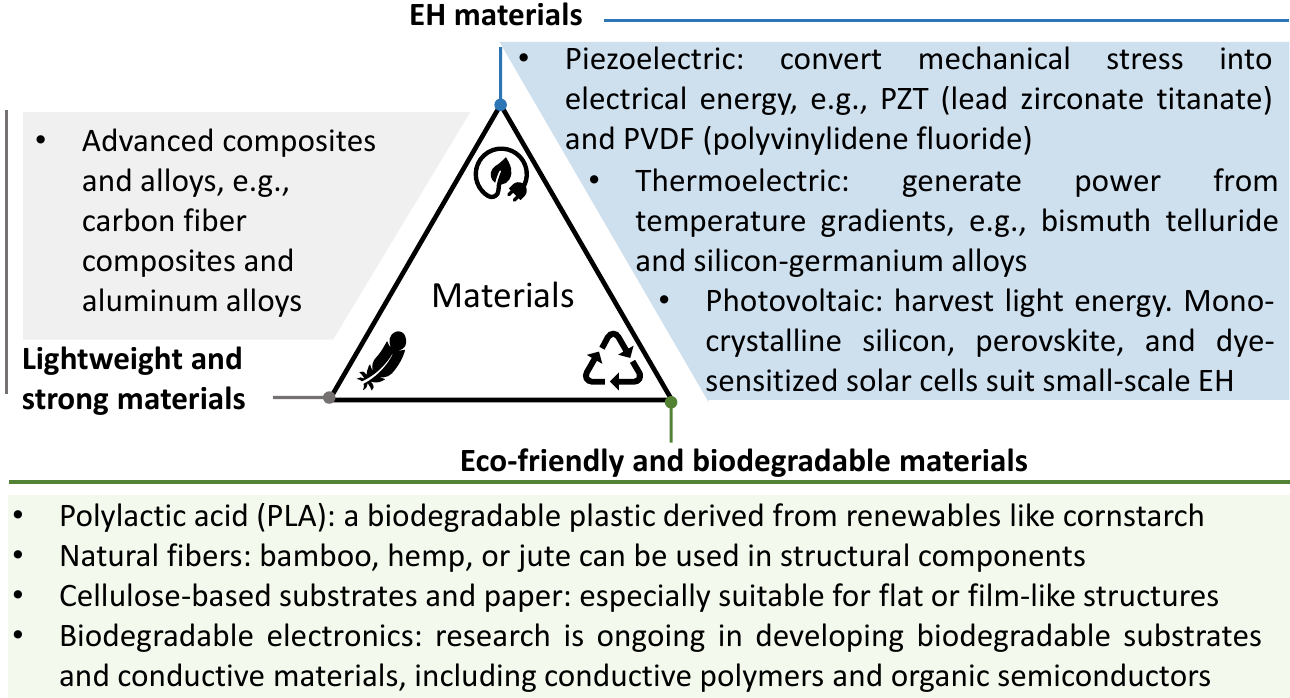}
    \caption{ZED material categories and some example choices.}
    \label{fig:materials}
\end{figure}

ZEDs may require incorporating EH materials depending on the EH source such as piezoelectric, thermoelectric, and photovoltaic materials. These and other ZED manufacturing materials should be, in most cases, biodegradable, low-cost,  lightweight, and robust. Fig.~\ref{fig:materials}  illustrates some example choices per category, but note that selecting materials that optimally balance these desirable properties often constitutes a complex engineering challenge. 

In terms of manufacturing, sustainable and circular practices must be adopted. These aim at minimizing waste and energy use by recycling and reusing materials. Also, additive manufacturing techniques such as printing, common in chipless RFID tags, may be further explored 
to mitigate the environmental challenges associated with the manufacturing, deployment, and disposal of other ZEDs types \cite{Wiklund.2021}. 
Such processes enable the creation of relatively complex, but lightweight, structures that would be difficult or impossible to make with traditional manufacturing.
Moreover, the additive process is often quicker and requires fewer materials, driving down both the capital and operational expenditures for manufacturing. This cost-effectiveness becomes particularly significant when manufacturing at scale. Notably, the lower production costs make it economically viable to produce backscatter devices for a wide array of applications where traditional (more expensive) devices are not feasible.
Finally, precision microfabrication, including photolithography, laser cutting, and micro-molding techniques, may be required for creating microscale energy harvesters and electronic components.

\subsection{Backscattering and low-complexity receivers} 
Backscattering constitutes a passive communication technique, thus suitable for ultra-low power ZEDs.
Indeed, a backscatter ZED modulates and reflects an incoming RF signal, rather than generating its own, i.e., no need for a local oscillator, thus requiring ultra-low power. 
The primary backscattering communication configurations are monostatic and bistatic \cite{Lopez.2023}. In a monostatic arrangement, a single antenna or a co-located pair serves dual roles, handling signals' transmission and reception. The reader emits an RF carrier signal, and the backscatter device modulates this signal by altering its antenna impedance before reflecting it for demodulation by the same antenna system. The co-existence of the RF source and the receiver means that both the transmit link and backscatter link distances are simultaneously increased when the backscatter device is moved away from the transceiver, leading to the doubly near-far problem.
Also, monostatic backscattering systems require rapid switching or some duplexing capability to operate effectively.\footnote{Note that full duplexing is not strictly needed but could be beneficial depending on the specific system requirements, including the need for simultaneous operation and the technical challenges associated with achieving sufficient isolation between transmit and receive functions.} 
Bistatic configurations, on the other hand, utilize separate antennas for the transmitting and receiving functions \cite{Liao.2023}, introducing more design flexibility at the expense of increased deployment complexity.

RF exciting signals may be dedicated or non-dedicated/ already-existing. The latter leads to the so-called ambient backscattering concept, eliminating the need for dedicated carrier emitters and additional frequency spectrum resources. Ambient backscattering is mostly adopted with bistatic configurations due to the inherent limitations imposed by the double near-far problem in such setups. We present a proof of concept related to ambient backscattering in Section~\ref{sec:poc}.


 Note that a backscatter device can easily implement 
 on-off keying (OOK) modulation by toggling between reflecting and absorbing the incident RF signals. When the device is reflecting, the modulated signal is sent back to the receiver, whereas, in the absorbing state, the signal is effectively nullified.
 Envelope detection receivers provide a simple, energy-efficient way to demodulate OOK backscatter signals.
 Specifically, the envelope detector captures the amplitude variations of the received RF signal, which directly correspond to the binary states of the OOK-modulated symbols. Because envelope detectors are relatively simple electronic circuits comprising components like diodes and capacitors, they require minimal processing power, are cost-effective, and thus may be also implemented in backscatter ZEDs for downlink communication, 
 enabling full-transceiver backscatter ZEDs.

\section{Protocols and Infrastructure Enablers \label{sec:enablers}}

Herein, we explore the pivotal protocol and infrastructure enablers for the seamless support and application expansion of ZEDs. Specifically, our discussions are geared toward highlighting the need for low-complexity communication, computation, (sleep/wake-up) scheduling, and sensing operations, potentially incorporating energy awareness and energy provision from the network infrastructure.

\subsection{Lightweight and energy-aware protocols} \label{protocols}

ZEDs cannot perform communication, computation, and/or sensing when the available energy is insufficient. Therefore, protocols for ZEDs must be adaptive and manage energy resources such that present and future system states are not compromised. This inevitably requires energy awareness and simplified layers and protocol designs \cite{Lopez.2023}.

Communication frame and slot patterns must be tweaked in 6G networks, e.g., to allow different slot, mini-slot, and frame design formats, including the possibility for the network to configure harvesting occasions. All these should adapt to the transfer data size, energy availability, and ZEDs' capabilities. 

At the medium access control layer, fast uplink grant and grant-free random access protocols are appealing due to their simplicity and low control signaling \cite{Mitsiou.2023}. Meanwhile, at the network layer, the RAN scope connection-oriented mode, e.g., the radio resource control connection-based approach in legacy cellular technologies, must evolve to a RAN scope dedicated connectionless communication. This leads to 
\emph{(i)} a limited 6G RAN connection, which does not involve dedicated connection handling and quality-of-service flow (no dedicated bearer provisioning), resulting in fewer handshakes between RAN nodes and ZEDs; 
and \emph{(ii)} a lightweight core network (CN) and user plane security provisioning. In general, this reduces monitoring load and signaling overhead.

ZED communication may be self-contained by incorporating a user RAN/CN identifier, which could be a short function of the CN identifier. 
%
%
There would be \emph{(i)} self-contained contention-based uplink, including an uplink preamble (optional) and synchronization signal, an uplink/downlink PHY header, and an uplink data block; and
    \emph{(ii)} self-contained downlink using a downlink paging signal containing device data.
Once the initial transmission is performed and the user is identified, the network can allocate short-lived device context and, if required, allocate additional resources for subsequent transmissions. 

Finally, system information required for RAN access and conditions/configurations qualifying a RAN node for device camping or paging is essential and, thus should be acquired immediately/frequently. Meanwhile, the acquisition of other non-essential information can be postponed, e.g., subject to network policy, ZED’s energy state, and next access attempts.




\subsection{Low-power machine learning}


TinyML constitutes computationally efficient and resource-limited machine learning (ML) algorithms tailored for ultra-low-power microcontrollers \cite{Sabovic.2023}. Its incorporation into ZEDs offers new on-device data processing and real-time prediction/decision capabilities, expanding their application horizons and/or boosting their KPIs. 
TinyML-equipped ZEDs may rely less on cloud connectivity and thus experience less communication latency, reduced transceiver energy consumption, and improved privacy.
Furthermore, TinyML models may be deployed to forecast future EH and energy expenditure, which in turn helps in devising proactive energy-aware/neutral operations for ZEDs.

A major challenge in deploying and running TinyML on ZEDs, are their limited capabilities and resources in terms of
\emph{(i)} memory, which limits the TinyML model size in the case of the non-volatile memory and the TinyML operation in the case of volatile memory;
\emph{(ii)} compute power (available energy), which steadily (instantaneously) limits the TinyML operation; and
\emph{(iii)} communication, which limits the TinyML capability to interact with edge/cloud nodes.

Table~\ref{tab:TinyML} lists potential TinyML algorithms with varying requirements in terms of memory and compute power during inference and their suitable applications. Note that unsupervised and reinforcement learning algorithms are preferred in general since such requirements are usually much larger during the training phase of supervised algorithms, except K-nearest neighbors (KNN). If the application strictly requires supervision, then the usual approach is to resort to offline training before deployment, thus limiting the dynamicity/adaptiveness/learning capacities of the host devices. This can be avoided if ZEDs report data to the cloud periodically thus triggering the eventual reception of TinyML model updates. What data and how often should be reported are critical issues to be explored as such a process may drain significant energy resources as well. Another approach is deploying large ML models at the edge and exploiting the federated learning framework such that several ZEDs collaboratively run TinyML models locally.

\begin{table*}[t!]
    \centering
    \caption{Potential TinyML base algorithms, their application tasks, and memory and compute power requirements during inference}
    \begin{tabular}{p{2.95cm}|p{4.5cm} p{4.65cm} p{4.6cm}}
    \toprule
         \textbf{TinyML algorithms}  & \textbf{learning application} & \textbf{non-volatile memory requirement}$^\dagger$ & \textbf{compute power requirement}   \\ \midrule
        \rowcolor{green!10}  low-order statistics (e.g., moving average, min/max, variance, count) &  lightweight online/sequential data processing, threshold-based policies \hfill $-$unsupervised & number of features & number of features \\ \hline       
        \rowcolor{green!10}  na\"ive Bayes & raw/low-order real-time classification \hfill $-$unsupervised &	number of features $\times$ number of classes &	number of features\\ \hline
        \rowcolor{green!10}  rule-based policies &	classification (which is soft in the case of fuzzy logic), decision-making, expert systems &	number of rules (plus number of linguistic terms and precision of the data representation in the case of fuzzy logic) &	number of rules (weighted by the number of membership functions for each linguistic variable in the case of fuzzy logic with Mamdani-type inference) \\ \hline
         \rowcolor{green!10} linear/logistic regression & regression, classification \hfill $-$supervised &	number of features $+$ 1 &	number of features \\ \hline
        \rowcolor{blue!10} Q-learning$^\ddagger$ & decision-making \hfill $-$reinforcement learning & number of features $\times$ number of states $\times$ number of actions &	number of features $+$ number of actions \\ \hline
        \rowcolor{blue!10} basic clustering (e.g., K-means) & classification, pattern recognition \hfill $-$unsupervised &	number of clusters $\times$ number of features per cluster (or data) &	number of clusters $\times$ number of features per cluster (or data) \\ \hline
        \rowcolor{blue!10} ARIMA &	univariate time series forecasting and raw anomaly detection \hfill $-$unsupervised &	number of autoregressive and moving average parameters	& number of autoregressive and moving average parameters \\ \hline
         \rowcolor{yellow!10} ensemble models, e.g., decision trees and random forests & regression, classification, anomaly detection, decision-making \hfill $-$supervised &	number of nodes	& number of nodes $\times \log$(number of features) \\ \hline
          \rowcolor{yellow!10} KNN (no training) & regression, classification, time series prediction \hfill $-$supervised	& dataset dimension $\times$ number of features &	dataset dimension $\times$ number of features \\ \hline
          \rowcolor{yellow!10} spectral and density-based clustering &	complex-shaped classification (e.g., for image segmentation, anomaly detection, shape recognition) \hfill $-$unsupervised &	number of features$^2$	& number of clusters $\times$ number of features$^2$ (although, it is usually lower for density-based clustering) \\ \hline         
         \rowcolor{red!10} support vector machines (SVM) & high-dimensional classification, regression, ranking, and anomaly detection \hfill $-$supervised, or unsupervised for one-class classification &	number of features $+$ number of support vectors $+$ other parameters if any (e.g., number of bias terms, slack variables, kernel matrix)	& number of features, in the case of linear SVM (for non-linear SVM, it also scales with the number of support vectors and the number of training instances depending on the adopted kernel) \\ \hline
        \rowcolor{red!10}  Gaussian mixture models &	classification (including anomaly detection and image segmentation), density estimation	\hfill $-$unsupervised & number of Gaussian components $+$ number of features &	number of Gaussian components $\times$ number of features$^2$ \\ \hline
        \rowcolor{red!10} (vanilla, but also deep) neural networks & regression, classification, function approximation, data compression, anomaly detection \hfill $-$supervised	 & number of neurons in the largest layer $\times$ (batch size (i.e., number of input samples) $+$ number of layers) $+$ number of features $+$ number of weights and biases &	batch size $+$ number of features $+$ number of weights and biases $+$ number of forward pass operations\\ \bottomrule
    \end{tabular}\\   
    \footnotesize{This is a general overview table, and the actual memory and compute power requirements can vary based on specific implementation details. The models are assumed to be already trained. Background colors qualitatively reflect their resource requirements, such that green, blue, yellow, and red represent low, mid-low, mid-high, and high levels, respectively.}\newline
     \raggedright{$^\dagger$\footnotesize{Volatile memory requirements scale similarly to the compute power requirement but including also the input data dimensions.}}  \newline
     \raggedright{$^\ddagger$\footnotesize{Volatile/non-volatile memory requirements of Q-learning models can scale quickly for a large number of features/states/actions, for which Q tables can be extremely large, thus difficult to store/update. In such cases, deep-Q networks, exploiting neural networks to approximate the Q-value function, are usually preferred and their resource requirements scale as shown in the last row.}}
    \label{tab:TinyML}
\end{table*}

In addition, there are several techniques to miniaturize ML models and their resource requirements, including \emph{(i)} architecture search to uncover the most suitable one; \emph{(ii)} parallel ultra-low power processors to provide software-level acceleration for TinyML models; \emph{(iii)} model compression techniques (e.g., quantization, pruning, and knowledge distillation), to reduce computational needs; and \emph{(iv)} dynamic random-access memory during inference \cite{Lopez.2023}. This does not come for free, and such techniques affect the models’ accuracy, evincing difficult-to-tame simplicity versus accuracy trade-offs. One approach for addressing this is to create several alternative TinyML models with different trade-off figures and then selecting one based on the instantaneous and foreseen availability of resources, especially energy. These alternative models can be generated offline, and stored in the ZED’s non-volatile memory. Alternatively, dynamic model execution approaches with runtime-adaptive energy consumption can be designed. Also, energy-aware execution of inference tasks ensuring enough energy is available for successful completion, e.g., based on worst-case power consumption and energy prediction models, can further alleviate power failures~\cite{Sabovic.2023}.

All in all, achieving optimal performance based on the hardware/software restrictions and capabilities is critical and this often requires customized designs/solutions. TinyML models operating efficiently, i.e., enhancing ZEDs' functionality and battery life, are certainly necessary to further support the decentralization of data processing capabilities, thus realizing scalable and cost-efficient intelligent ecosystems.

\subsection{RF-WPT}

WPT may be pivotal in supporting the widespread deployment and operation of ZEDs by providing a controllable/predictable energy supply. Among the diverse WPT technologies, RF-WPT is particularly appealing for this \cite{Naser.2023,Rosabal.2023}, mainly due to its inherent capability of broadcasting energy over long distances and charging multiple devices simultaneously even in non-line-of-sight conditions. Moreover, RF-EH circuit form factors and manufacturing costs allow seamless integration of this technology in existing devices and enables dual EH from both dedicated and ambient energy sources. 

Next-generation base stations (BSs) may be designed and configured to support WPT in addition to legacy and enhanced data transmission services. In this way, ultra-low power ZEDs may be wirelessly charged with predefined performance guarantees. However, since the charging efficiency decreases exponentially with the charging distance, this might be only possible in highly dense BS deployments, e.g., ultra-dense urban areas and private indoor networks. In other scenarios, the deployment of dedicated WPT nodes, usually referred to as power beacons (PBs) \cite{Rosabal.2023}, is needed. Although this entails higher upfront costs, overall network costs can be reduced compared with traditional IoT setups, especially as the number of ZEDs increases and considering that inaccurate ZED power profiling, battery imperfections, and/or operating conditions can shorten battery lifespans far below expected \cite{Rosabal.2023}. Meanwhile, overall costs may be similar to those of ZED deployments exploiting other EH solutions, but this requires further study.

The adoption of low-power/cost multi-antenna architectures at the PBs is needed to increase charging efficiency and promote economic feasibility \cite{Lopez.2023}. That is the case with dynamic metasurface antennas and radio stripe networks. The former is a novel hybrid multi-antenna technology consisting of several waveguides, where each one is connected to a dedicated RF chain and feeds multiple radiating metamaterial elements. Meanwhile, in the latter, several radiating elements are implemented along a cable with one or more central processing units while their implementation complexity is independent of the number of elements thanks to their compute and forward architecture. The integration of such technologies in future wireless networks, including for supporting RF-WPT, still calls for lightweight and efficient dedicated protocols.

\subsection{Duty Cycling and Wake-up Protocols}\label{WU}

ZEDs may need to sleep for long periods to save or harvest enough energy for executing their relevant computation, sensing, and/or communication tasks \cite{Naser.2023}. Therefore, their uplink and downlink transmission and active and sleep times, i.e., duty cycling, must adapt to energy availability, which hereinafter also includes EH capabilities, battery state, and the possibility of receiving energy through WPT. 

ZEDs and network nodes must tightly cooperate for optimized performance. Specifically, ZEDs must provide energy-availability information while network nodes and ZEDs must adjust their listening intervals and data transmissions.  The network nodes can also exploit such information to properly schedule the transmission of wake-up signals (WuS) to active ZEDs.\footnote{WuS were already adopted in 3GPP Release 15 as downlink PHY signals before paging and enhanced in 3GPP Releases 16-18 with cross-slot scheduling, group-based wake-ups, and novel PHY and higher layer procedures.}
Note that information such as harvesting capabilities and storage size should be provided at the registration time through the indication of the device type or device category, while information related to traffic conditions, amount of data, energy storage level, and instantaneous energy sources availability may be provided frequently, e.g., during the actual data transmission.

ZEDs must incorporate an ultra-low power wake-up radio (WuR) for the implementation of network-triggered wake-up signaling 
such that the main radio, if any, is only activated for data transmission and reception. In the case that no other radio is available, the WuR enters a data reception/transmission stage. 
Note that in traditional IoT setups, WuRs' average power consumption can be $\sim30$~dB lower than that of the main radio, while 
wake-up signaling implementations perform traditionally better than duty cycling protocols under light traffic load~\cite{Magno.2016}. In WuR-equipped ZEDs, however, the power consumption gap between the main radio, if any, and WuR may not be that significant, so these figures and trends need to be re-assessed. 
Challenging aspects related to the densification, stringent energy limitations, and application KPIs of future ZED networks must be taken care of. This includes taming the traditional WuR sensitivity and selectiveness trade-offs but tuned to ZED particularities. 
When monitoring for a WuS, high sensitivity and selectiveness are desired to avoid miss-detection and false alarm errors, respectively. While the former incurs extra network resources and delay, the latter results in unnecessary energy consumption at the ZED.
WuR may be highly advantageous if optimized for always-on battery-less listening, supported, e.g., by ambient RF EH, avoiding WuR duty cycling and corresponding sensitivity degradation.
Meanwhile, the implementation of a WuR is particularly challenging in emerging massive multi-antenna networks exploiting high frequencies due to energy and signaling overhead limitations affecting beam sweeping procedures \cite{Lopez.2023}. All these issues and the need for in-band WuR implementations complicate radio resource management and scheduling.

Noteworthy, ultra-low power ZEDs may not incorporate a local oscillator and in such cases, their duty cycling may differ significantly from conventional timing-based systems. In such a case, the sleep and wake-up periods may be configured entirely based on energy availability rules. In general,
duty cycling and wake-up protocols can benefit from on-device and edge/network intelligence, respectively. Indeed, ZEDs could incorporate TinyML models to schedule sleeping periods based on energy availability, thus facilitating autonomous operation, and eliminating unnecessary communication loops during idle periods. On the other hand, edge/network nodes could leverage advanced ML-based algorithms for that purpose, e.g., by learning traffic profiles and energy availability and power consumption dynamics \cite{Guirola.2022}, thus lightening the computation/prediction-related tasks at the ZEDs. As usual, there is no one-size-fits-all solution, and properly tuned hybrid approaches may be preferred.

\section{Cellular-backscattered ZEDs:\\ A Proof of Concept \label{sec:poc}}

\begin{figure}
    \centering
    \includegraphics[width=0.45\textwidth]{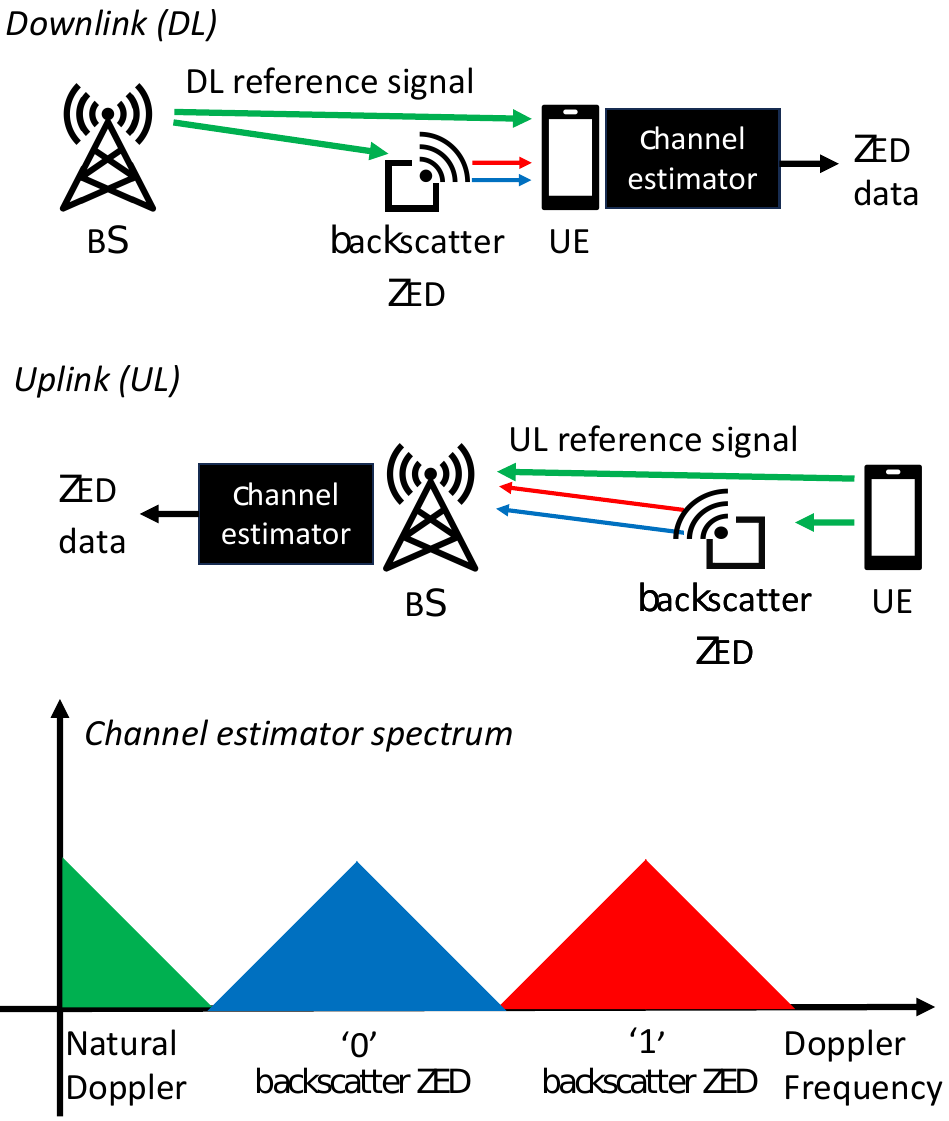}
    \caption{Using cellular infrastructure to read backscatter modulated messages. A backscatter ZED introduces a frequency shift to the scattered signal. A channel estimator can separate the natural Doppler components from the Backscatter ones induced in the Doppler frequency domain.}
    \label{fig:ZED_PoC}
\end{figure}

 Backscatter ZEDs can be integrated into existing mobile communication systems. Note that a backscattered signal appears as an additional multipath component at a receiver, but with an artificially induced Doppler shift owing to its modulation. Therefore, existing channel estimators
can decode the backscatter-modulated downlink or uplink signals as long as the corresponding artificial Doppler shifts are within the maximum tolerable Doppler limits of the system. Notably, mobile communication systems are generally designed to operate even in high-velocity environments, while the mobility level in the realm of low-power applications is typically more moderate. This leaves available bandwidth in the channel estimator, thereby creating an opportunity to accommodate the signals from backscatter ZEDs without disrupting the existing system's performance. Fig.~\ref{fig:ZED_PoC} illustrates the idea.

\begin{figure}
    \centering
    \includegraphics[width=0.4\textwidth]{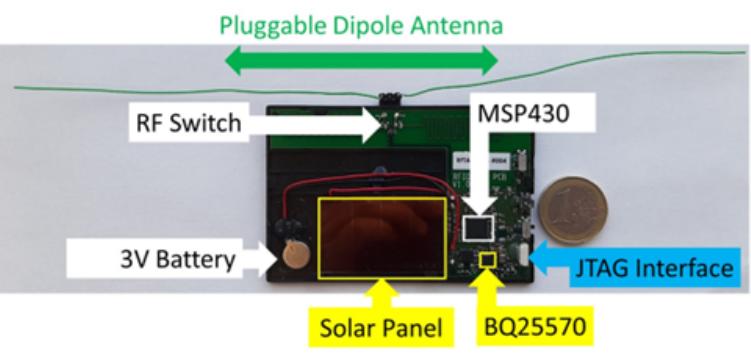}
    \caption{ZED prototype consuming around 50 $\mu$W to continuously transmit $\sim$100 bits every 10s. 
    }
     \label{fig:prototype}
\end{figure}

The concept of seamlessly integrating backscatter ZEDs into existing mobile communication systems has been empirically demonstrated in the context of LTE downlink transmissions in \cite{Liao.2023}. Therein, a software-defined radio implementation emulates the user equipment (UE) channel estimator, showcasing that it is indeed possible to successfully decode backscatter-modulated signals. The experimental results confirm that ZEDs' detection by LTE smartphones does not require hardware changes. The fluctuations induced by ZEDs on the network-to-smartphone links are slow enough to be tracked by smartphones, which are standardized to track faster fluctuations (e.g., in high-speed trains). By enabling smartphones to interpret fluctuations that they already track, we pave the way for the smooth integration of ZEDs into existing networks. This system has recently been tested with a commercial LTE network as an ambient source and a ZED prototype ~\cite{Phan.2022}  illustrated in  Fig.~\ref{fig:prototype}.

For the ZED prototype to transmit bit '0' and bit '1', the two branches of a dipole antenna are connected and disconnected with a period T0 and T1, respectively, thanks to an RF switch.  When the branches are connected, the prototype backscatters ambient waves, while the prototype is transparent to waves when they are disconnected. The prototype does not consume RF-wave generation power, while a low-power Texas Instrument MSP430 microcontroller controls the RF switch to send a fixed and periodical message. An ultra-low power harvester power management integrated circuit BQ25570 from Texas Instruments manages \emph{(i)} the EH of indoor and outdoor light by a solar panel, \emph{(ii)} the energy storage in a 3V rechargeable battery, and \emph{(iii)} the circuit energy consumption. Analysis of the energy consumption shows that the prototype consumes around 50 $\mu$W to continuously transmit $\sim$100 bits every 10s and can thus operate infinitely and autonomously, provided $\sim 10$ hours per day of bright light (corresponding to $\sim$150~$\mu$W harvesting power).

\section{Conclusion \label{sec:conclusion}}
This paper identified the key characteristics of ZEDs together with their latest advancements and technological enablers, including techniques, protocols, and infrastructure elements.
Specifically, we discussed suitable unconventional EH sources, multi-source EH and power management techniques, energy storage solutions, manufacturing materials and practices, and backscattering communication systems including low-complexity receivers, and highlighted the need for lightweight and energy-aware computing, communication, and scheduling protocols. Low-complexity computing/intelligence mechanisms for ZEDs were thoroughly discussed, together with duty cycling, and infrastructure enablers like radio frequency wireless power transfer and wake-up protocols. We revealed related challenging aspects and open research directions, and showcased an experimental ZED proof-of-concept related to ambient cellular backscattering. 
This research opens new avenues for the development of sustainable, energy-efficient IoT devices, paving the way for a more connected and environmentally conscious future. The insights garnered from this study are instrumental for researchers and practitioners in the field, providing a roadmap for future advancements in the realm of ZEDs and their applications in various sectors.

\section*{Acknowledgments}
This work has been partly funded by the European Union’s Horizon Europe research and innovation programme under grant agreement No. 101095759 (Hexa-X-II) and by the Research Council of Finland (former Academy of Finland) 6G Flagship Programme (Grant Number: 346208).


\bibliographystyle{IEEEtran}
\bibliography{references}

\end{document}